\documentclass[
	paper=a4,
	DIV=27,+-++++
	fontsize=11pt	
	]{scrartcl}

	\KOMAoption{captions}{tableheading, figuresignature, nooneline, bottombeside, outerbeside}
	\addtokomafont{caption}{\small}			% Schriftgröße für die Captionsfestlegen
	\addtokomafont{captionlabel}{\bfseries} % Einstellung für das Captionlabel
	
		 \addtokomafont{author}{\small}
		 \addtokomafont{date}{\small}
		 \addtokomafont{publishers}{\small}
\usepackage[utf8]{inputenc} 
\usepackage[T1]{fontenc}
\usepackage[english]{babel}
\usepackage{ulem} %Text durchstreichen
\usepackage[sfdefault, light, condensed]{roboto}	    % TUC Corporate Design
\usepackage{amsmath}				% Einbinden von Mathematik
\usepackage{amssymb}				% Einbinden von mathematischen Symbolen
\usepackage{siunitx}	            %SI-Einheiten in "richtiger" Darstellung
	\DeclareSIUnit{\angstrom}{\textup{\AA}}

\usepackage[version=4]{mhchem}      % Einbinden von chemischen Formeln

\usepackage[onehalfspacing]{setspace} % Setzen des Zeilenabstande
\usepackage{graphicx}
	\graphicspath{ {images/} }

\usepackage{authblk}	%Author and Affilliation
\usepackage{textgreek}	%griechische Buchstaben im Fließtext
    \usepackage{upgreek}

\usepackage[dvipsnames]{xcolor}

\usepackage{hyperref}

\usepackage[sort&compress,square,numbers,comma]{natbib}

\title{Si Intercalation Beneath Epitaxial Graphene: Modulating Mott States at the SiC(0001) Interface}
	\author[1,2]{Niclas Tilgner}
    \author[1]{Zamin Mamiyev}
	\author[1,2]{Susanne Wolff}
    \author[1,2]{Philip Sch\"{a}dlich}
    \author[1,2]{Fabian G\"{o}hler}
    \author[1]{Christoph Tegenkamp}
	\author[1,2,*]{Thomas Seyller}
	
	\affil[1]{Institute of Physics, Chemnitz University of Technology, 09126 Chemnitz, Germany}
	\affil[2]{Center for Materials, Architectures and Integration of Nanomembranes (MAIN), 09126 Chemnitz, Germany}
	\affil[*]{\textit{Corresponding author E-Mail:} thomas.seyller@physik.tu-chemnitz.de}

\begin{document}

%=================
%	Titelseite
%=================	
	
	\maketitle
	
	 \par
	
	\begin{abstract}

    Intercalation has proven to be a powerful tool for tailoring the electronic properties of freestanding graphene layers as well as for stabilizing the intercalated material in a two-dimensional configuration.
    This work examines Si intercalation of epitaxial graphene on SiC(0001) using three preparation methods. Dangling bond states at the interface were found to undergo a Mott-Hubbard metal-insulator transition as a result of a significant on-site repulsion.
    Comparing this heterostructure consisting of graphene and a Mott insulator with a similar system without graphene, reveals the screening ability of graphene's conduction electrons on the on-site repulsion.
    The system presented here can serve as a template for further research on Mott insulators with variable band gap.
	
	\end{abstract}

%=================
%	1 - Introduction 
%=================	

    \section{Introduction}
	
    Graphene, a two-dimensional (2D) allotrope of carbon, where the atoms arrange in a honeycomb structure, has developed since its first experimental realization \cite{novoselov_geim} as a viable option for use in energy storage \cite{graphene_energy_storage, graphene_composites}, material science \cite{graphene_composites, anti-corrosive_coatings}, electronics \cite{graphene_transistors}, and other fields. The crucial properties for those applications are the high charge carrier mobility \cite{novoselov_geim} and thermal conductivity \cite{thermal_conductivity}, as well as the mechanical strength and flexibility \cite{flexibility}. Consequently, preparing and tailoring graphene is a primary area of interest in contemporary research.

    One established preparation technique is the epitaxial growth on SiC(0001) \cite{de2007epitaxial, Emtsev_6r3_auf_SiC, Sublimationswachstum, PASG}, where the initial graphene layer grown in this manner (commonly known as buffer layer or zero layer graphene, ZLG) is hybridized with the substrate's surface dangling bonds, i.e., it lacks the properties mentioned before \cite{Emtsev_6r3_auf_SiC}.
    Intercalation of foreign atoms between ZLG and substrate provides a tool not only for decoupling the graphene \cite{H_Riedl, H_Interkalation}, but also for sculpting heterostacks with other unusual 2D materials. Examples are topological insulators \cite{schmitt_indenene, tilgner2025reversible}, Mott insulators \cite{RuCl3_1, RuCl3_2}, and superconductors \cite{SC_Li, SC_Bernado}. Examining whether such extraordinary characteristics transfer to the graphene just by virtue of proximity is an interesting question.

    Particular attention was attracted in the past by group-IV adatoms, i.e. C, Si, Ge, Sn, and Pb, on hexagonal substrates with a 1/3 monolayer coverage and a ($\surd{3} \times \surd{3}$)$R30^{\circ}$ periodicity \cite{ganz1991submonolayer, profeta2005novel, profeta2007triangular, glass2015triangular, Johansson_r3}. These lattice systems exhibit dangling bond states, which undergo a Mott-Hubbard metal-insulator transition, thus, showing insulating behavior although being half-filled as a result of a significant on-site repulsion \cite{Hubbard, Hubbard3, Mott}. Very recently, theoretical considerations predicted a strong hybridization of these 2D group-IV Mott insulators when brought into proximity with graphene \cite{witt2025quantum}. The magnitude of this hybridization was found to depend on the size of the adatoms, with C and Si showing stronger couplings than Sn and Pb. In all cases, however, the hybridization led to correlated flat-band physics, which may open up interesting quantum phases such as superconductivity, magnetism, or topological states \cite{witt2025quantum}. Experimentally, a recent study \cite{ghosal2024electronic} has shown that the realization of a Sn layer with ($\surd{3} \times \surd{3}$)$R30^{\circ}$ symmetry at the graphene-SiC interface does indeed lead to a moderate hybridization with graphene.
    
    The present work investigates the intercalation of Si, which is known to form a ($\surd{3} \times \surd{3}$)$R30^{\circ}$ \cite{starke_T4_distance} as well as a ($3 \times 3$) \cite{schardt_3x3} reconstruction on the SiC(0001) surface. Both phases exhibit dangling bond states, which were shown to undergo a Mott-Hubbard metal-insulator transition \cite{Johansson_r3, furthmüller_3x3}. Although there have been a few studies on Si intercalation of epitaxial graphene in the past \cite{Xia, visikovskiy2016graphene, silly2014electronic1, silly2014electronic2}, none of them reported the formation of a Mott insulator at the graphene-SiC interface.
    
    This study focuses on three preparation methods: (A) sequential deposition and annealing, (B) deposition at elevated temperatures and (C) exchange intercalation, where Si is intended to replace H in a sample previously intercalated by hydrogen. The approaches are compared in terms of the effectiveness of the intercalation process, the graphene doping, and the surface quality. The first two approaches lead to the occurrence of a surface state attributed to the anticipated lower Mott-Hubbard band of Si dangling bonds at the interface. Subsequent comparison with the uncovered analogs demonstrates the screening ability of graphene's conduction electrons, which reduce the on-site repulsion of the dangling bonds, thereby shrinking the band gap of the Mott insulator.
    Contrary to the theoretical considerations mentioned above \cite{witt2025quantum}, which predict a strong hybridization between the Si dangling bonds and graphene, no evidence of such is found, since the band structure of the Si intercalated graphene remains undisturbed.

%=================
%	2 - Results & Discussion
%=================	
    
\section{Pathways towards Si Intercalation}\label{sec_XPS}

    \begin{figure}[b!]
        \centering
        \includegraphics[scale = 1]{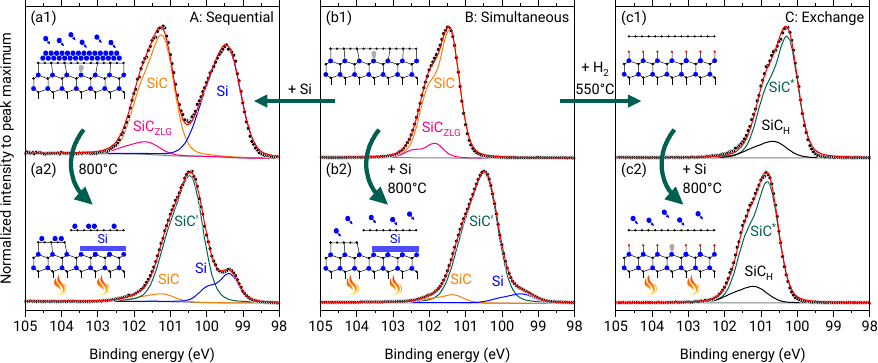}
        \caption{X-ray photoelectron spectra of the Si\,2p core level recorded after different preparation steps, all starting from an epitaxially grown zero layer graphene (ZLG) sample. A representative spectrum for the ZLG is shown in (b1), featuring contributions from bulk Si (SiC) and topmost Si (SiC$_{\text{ZLG}}$) atoms coupled to the ZLG.
        Method A: Sequential. (a1) Si was deposited onto the ZLG, resulting in an additional component labeled Si. (a2) Annealing the sample to \qty{800}{\degreeCelsius} caused significant spectral changes, modeled with a new component (SiC'), indicating modification of the interface.
        Method B: Simultaneous. (b2) Si deposition and annealing were performed simultaneously, producing a spectrum similar to (a2), with slightly different contributions.
        Method C: Exchange. (c1) The ZLG was decoupled via H intercalation by annealing in a hydrogen environment, yielding a spectrum with a shifted bulk contribution (SiC$^{*}$) and a new interface component (SiC$_{\text{H}}$), reflecting hydrogenation of the graphene-SiC interface. (c2) Subsequent Si deposition and annealing led to an energy shift in the spectrum.
        The insets show ball-and-stick models illustrating the individual preparation steps, with Si, C, and H atoms color-coded in blue, black, and red, respectively.
        }
        \label{Si2p}
    \end{figure}

    The starting point of all three preparation methods was an epitaxially grown ZLG sample on n-type 4H-SiC(0001). A representative X-ray photoelectron (XP) measurement of the Si\,2p core level is presented in \autoref{Si2p}\,(b1). The data was fitted with two components corresponding to bulk and topmost Si atoms. Note that each component discussed in this section consists of two Voigt peaks with an energy splitting of \qty{0.6}{eV} and an intensity ratio of 2:1, taking into account the spin-orbit splitting of the Si\,2p orbital \cite{Diss_Emtsev}. The contribution of the bulk atoms is represented by the component with highest intensity labeled SiC. The binding energy of the Si\,2p$_{\text{3/2}}$ electrons is \qty{101.5}{eV}, which is equivalent to an upwards band bending of \qty{0.4}{eV} \cite{mammadov}. This value is fixed by Fermi level pinning, because of surface dangling bonds \cite{Emtsev_6r3_auf_SiC}. Furthermore, the chemically shifted component $\text{SiC}_\text{ZLG}$ takes the Si atoms into account, that are covalently bound to the ZLG. The chemical shift was constrained to \qty{0.4}{eV}, according to Ref. \cite{Diss_Emtsev}.

    For the first preparation technique denoted A, 7 to 9 layers of Si were deposited onto the ZLG surface, giving rise to the additional component labeled Si in \autoref{Si2p}\,(a1), that is located at \qty{99.4}{eV}. The rest of the spectrum remains unchanged, implying that the interface has not been altered. Afterwards, the sample was annealed at \qty{800}{\degreeCelsius} for \qty{30}{min}, which led to several changes in the spectrum, as can be seen in \autoref{Si2p}\,(a2). It becomes apparent that a new component SiC', which is likewise allocated to the substrate, had to be added. The difference in binding energy between SiC and SiC' is 0.8\,eV, suggesting that the band bending became stronger and therefore the interface was modified. This is attributed to the intercalation of Si atoms between ZLG and substrate.
    Further evidence for a successful intercalation is provided by the corresponding C\,1s XP spectra discussed in Section 1 of the Supplementary Material, which show the same evolution of the bulk component and, most importantly, the replacement of the ZLG components by a single asymmetric peak attributed to the metallic quasi-freestanding monolayer of graphene (QFMLG).
    Note also that both bulk species (SiC and SiC') are present simultaneously, indicating an incomplete (\qty{90}{\%}) intercalation. Furthermore, the Si component was reduced in intensity, which leads to the conclusion that most of the Si desorbed from the surface during annealing. The binding energy from the Si component remained the same, since the bonding partners are solely Si atoms both in the deposited layer and at the interface.

    In the second preparation technique labeled B, the same steps as above were executed, but simultaneously. The ZLG was heated to \qty{800}{\degreeCelsius}, and approximately 1 layer of Si was deposited onto the surface. This process took \qty{30}{min}. The resulting Si\,2p core level spectrum, which closely resembles the spectrum discussed previously, is displayed in \autoref{Si2p}\,(b2). It includes the intercalated bulk as well as a comparably small fraction of not-intercalated areas with the same shift as for method A. From this, it is concluded that the interface has undergone a similar transformation. The only difference is found in the Si component, which has an appreciable lower intensity than in \autoref{Si2p}\,(a2). Further preparation steps could not increase the intensity. Consequently, it is assumed that while most of the Si is intercalated by technique B, a significant quantity of residual Si rests on top of the graphene after applying method A.

    For the third preparation technique denoted C, the ZLG was first decoupled with H as it is described in Ref. \cite{H_Riedl}. The Si\,2p core level spectrum at this stage is displayed in \autoref{Si2p}\,(c1). Two components were used for the fit. In analogy with the earlier findings, SiC$^{*}$ corresponds to the Si bulk atoms underneath the regions intercalated with H.
    As described by Ristein \cite{Ristein}, Mammadov \cite{mammadov}, and co-authors, the band bending is then controlled by the spontaneous polarization, which is inherent to the pyroelectric substrate. In the present case the binding energy of SiC$^{*}$ is \qty{100.3}{eV}, which corresponds to a band bending of \qty{1.6}{eV}. Furthermore, the component $\text{SiC}_\text{H}$ shifted by \qty{0.3}{eV} takes the Si-H bonds at the surface into consideration \cite{H_Riedl}.

    Analogous to the previously described preparation method B, the sample was subsequently annealed at \qty{800}{\degreeCelsius} and simultaneously exposed to a Si flux for \qty{30}{min}.
    \autoref{Si2p}\,(c2) shows the XP measurement of the Si\,2p core level after this step, which was fitted analogous to \autoref{Si2p}\,(c1). Interestingly, compared to \autoref{Si2p}\,(c1) the whole spectrum is shifted by \qty{0.5}{eV} towards higher binding energies, indicating a change in band bending and thus implying a modification of the interface. As can be seen below (cf. \autoref{cones_UPS}\,(c2)), the Dirac cone is present at the graphene $\text{K}_\text{G}$ point, indicating an intact graphene layer, and hence suggesting that the shift is not the result of large-scale deintercalation. However, the shoulder in the Si\,2p core level spectrum, previously attributed to intercalated Si, is absent, suggesting that Si is not adsorbed either on the graphene or at the SiC interface. The energy shift is assumed to be the result of sporadic H desorption, which is known to occur in this temperature regime \cite{H_Riedl}, resulting in isolated dangling bonds that act as electron reservoirs. This reduces the space charge at the interface and hence the band bending compared to the initial H intercalated state.
    
    Apparently, the H desorption is too marginal for Si intercalation at the stage presented here, so that it is simply blocked. However, upon further processing of this sample, i.e. further H desorption, Si intercalation was also observed for this preparation approach.
    Please refer to Section 2 of the Supplementary Material for a discussion of the evolution of the Si\,2p core level spectrum as the Si processing of this sample continues.

\section{Graphene Doping and Mott-Hubbard States}

    \begin{figure}[b!]
        \centering
        \includegraphics[scale = 1]{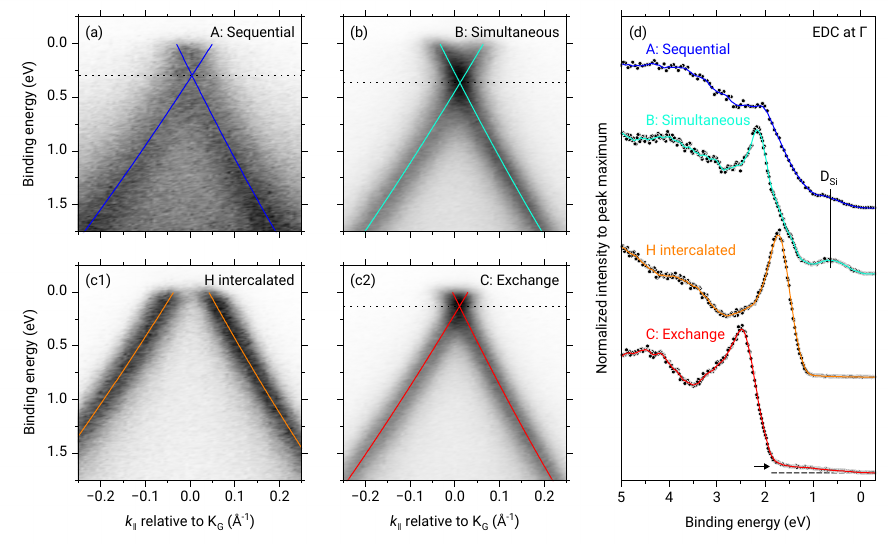}
        \caption{(a)-(c) Energy-momentum cuts in the vicinity of the graphene $\text{K}_\text{G}$ point perpendicular to $\overline{\text{\textGamma}\text{K}_\text{G}}$ after different preparation steps. A Dirac cone can be observed in every panel, demonstrating the effectiveness of each intercalation approach.
        The straight lines depict the fitted band dispersions. The dashed lines indicate the Dirac energies.
        (d) Corresponding energy distribution curves (EDC's) at $\text{\textGamma}$. The strong contributions observed from \qty{1.0}{eV} towards higher binding energies are ascribed to the SiC valence bands. Only the preparation techniques A and B lead to the additional appearance of the surface state denoted $\text{D}_\text{Si}$ at \qty{0.6}{eV}, which is attributed to the lower Hubbard band of the Mott insulator that formed at the interface.
        }
        \label{cones_UPS}
    \end{figure}

    \autoref{cones_UPS}\,(a)-(c) display energy-momentum cuts in the vicinity of the $\text{K}_\text{G}$ point of graphene after the different preparation steps described above. All demonstrated methods result in a linear dispersion at $\text{K}_\text{G}$, assigned to the Dirac cone of a QFMLG \cite{first_cone}. This demonstrates that the ZLG can be successfully decoupled using any of the three approaches described above.

    To determine the Dirac energy and, consequently, the doping of the QFMLG, a tight-binding fit was applied to the data. For this purpose, the dispersion of the \textpi-band was estimated by fitting momentum distribution curves with a Lorentzian. The extracted maxima were then fitted with the nearest-neighbor tight-binding band structure of graphene. The value of the overlap integral was constrained to 0.129 \cite{Saito}, whereas the exchange integral as well as the Dirac energy were estimated from the fit. For all samples, the Fermi velocity was found to be consistent with the theoretical value of \qty{1.1e6}{m/s} \cite{Saito}.

    Techniques A (\autoref{cones_UPS}\,(a)) and B (\autoref{cones_UPS}\,(b)) lead to a similar graphene doping (Dirac energy), which is a further indication that the interface between SiC and graphene is comparable in both approaches. The charge carrier density of the n-doped graphene in the two cases ranges from \qty{5e12}{cm^{-2}} to \qty{8e12}{cm^{-2}}. A difference is found in the signal-to-noise ratio, that is much lower in \autoref{cones_UPS}\,(a) than in \autoref{cones_UPS}\,(b). This may be explained by residual Si sitting atop the graphene, which increases the probability of inelastic or diffuse elastic scattering of the photoelectrons. From the XP spectra discussed before, it was already suspected that there might be residual Si at the surface.

    \autoref{cones_UPS}\,(c1) shows the Dirac cone of the QFMLG intercalated solely with H, which is p-doped with a hole density of \qty{5e12}{cm^{-2}} as already reported by previous studies \cite{H_Interkalation, mammadov}. After the process with Si in \autoref{cones_UPS}\,(c2), the doping changes to n-type with an electron density of \qty{1e12}{cm^{-2}}. In the context of \autoref{Si2p}\,(c2), a lower band bending was observed, indicating a reduction of the space charge at the SiC interface compared to the initial H intercalated stage, which is attributed to isolated dangling bonds acting as an electron reservoir.
    Such a reservoir can also explain the electron doping of the QFMLG after the Si process as observed in \autoref{cones_UPS}\,(c2).

    Energy distribution curves (EDC's) at $\text{\textGamma}$ for each of the four intercalated phases are displayed in \autoref{cones_UPS}\,(d). Strong contributions from \qty{1.0}{eV} to higher binding energies are present in all spectra and are ascribed to the SiC valence bands. According to the XP spectra (cf. \autoref{sec_XPS}) the various approaches have different band bendings. The valence band's relative shifts reflect the same trend as the core levels.

    Furthermore, the EDC B shows an additional contribution labeled $\text{D}_\text{Si}$ at \qty{0.6}{eV} within the band gap. This feature is also present in the curve A, but with a worse signal-to-noise ratio that can be again linked to the residual Si at the top. It is assumed, that the observed state $\text{D}_\text{Si}$ is the LHB of Si dangling bonds at the interface. Si rich ($\surd{3} \times \surd{3}$)$R30^{\circ}$ and ($3 \times 3$) reconstructions of the bare SiC(0001) surface are known to possess half-filled Si dangling bonds, where the electron-electron interaction is stronger than their delocalization \cite{Johansson_r3, Northrup_2, furthmüller_3x3}. Consequently, a Mott-Hubbard metal-insulator transition \cite{Hubbard, Mott} gives rise to a fully occupied lower Hubbard band (LHB) and an empty upper Hubbard band (UHB). Note that both are frequently referred to as Mott states \cite{witt2025quantum, ghosal2024electronic}.

    In accordance, the H intercalated sample does not show this surface state, since all dangling bonds are saturated. Furthermore, no clear peak is observed in the bulk band gap of EDC C, supporting the previous assumption that no Si atoms are intercalated. However, in contrast to the pure H intercalated sample, EDC C shows an increased background within the band gap, which is highlighted with an arrow in \autoref{cones_UPS}\,(d). This is due to the aforementioned isolated dangling bonds that act as pinning centers for the substrate and the graphene, but are too few to make a peak-like contribution.
    Upon further processing of this sample (see Section 2 of the Supplementary Material), Si intercalation occurs along with a more peak-like feature attributed to the LHB, as observed for the other two preparation approaches.

    A previous study of Si intercalation on SiC(0001) by Xia et al. \cite{Xia} detected a surface state at \qty{1.5}{eV} only after annealing at \qty{1000}{\degreeCelsius}. They assume that the Si reacted with the graphene and formed SiC with an additional ($\surd{3} \times \surd{3}$)$R30^{\circ}$ reconstruction on top. Earlier investigations of the unreconstructed \cite{emtsev_correlation_effects} and ($\surd{3} \times \surd{3}$)$R30^{\circ}$ reconstructed \cite{Johansson_r3} SiC(0001) surface also reported a state in this energy regime, which is assigned to the LHB of the Si dangling bonds. Nevertheless, a large area formation of SiC is excluded here because the preparation temperature was significantly lower than \qty{1000}{\degreeCelsius} and the Dirac cone is clearly visible (cf. \autoref{cones_UPS}\,(b)), suggesting an intact graphene layer. Consequently, it is proposed that Si forms a Mott insulator underneath graphene. The on-site repulsion is reduced by the free charge carriers of the metallic QFMLG. Thus, the decreased binding energy of \qty{0.6}{eV} compared to Xia et al. \cite{Xia} can be explained by screening of the on-site repulsion \cite{Hubbard}. A similar observation was recently \cite{ghosal2024electronic} made for a ($\surd{3} \times \surd{3}$)$R30^{\circ}$ reconstructed Sn layer at the graphene-SiC interface, where the LHB of Sn dangling bonds was found closer to the Fermi level compared to the analogous structure without graphene \cite{glass2015triangular}.

    In contrast to recent theoretical considerations \cite{witt2025quantum} of the Si rich ($\surd{3} \times \surd{3}$)$R30^{\circ}$ reconstruction on SiC(0001) in the proximity of graphene, no hybridization was observed in the systems presented here. The Dirac cone of graphene was found to be unperturbed, thus excluding strong interactions with neighboring materials. A contributing factor to this discrepancy could be the incommensurability of graphene with the substrate and consequently its reconstructions, leading to many different relative arrangements of the Si dangling bonds and the graphene C atoms, and consequently different hybridization strengths. However, the authors of Ref. \cite{witt2025quantum} only considered the case where the C atoms are directly above the dangling bonds.

\section{Electronic and Geometric Structure of Si at the Graphene-SiC Interface}

    \begin{figure}[b!]
        \centering
        \includegraphics[scale = 1]{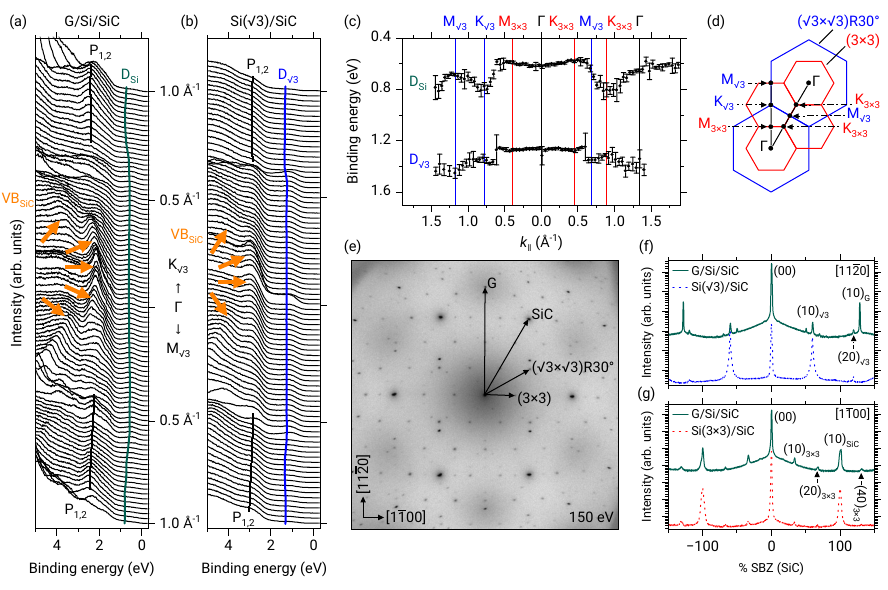}
        \caption{Electronic and geometric structure of Si at the interface between graphene and SiC.
        (a)-(b) Energy distribution curves along the $\overline{\text{K}_{\surd{3}}\text{\textGamma}\text{M}_{\surd{3}}}$ direction at different $k_\|$ values for Si intercalated graphene (G/Si/SiC) (a) and the Si rich ($\surd{3} \times \surd{3}$)$R30^{\circ}$ reconstruction without graphene (Si($\surd{3}$)/SiC) (b), both on 4H-SiC(0001). Contributions from the substrate valence band ($\text{VB}_{\text{SiC}}$) are highlighted with orange arrows. The features marked by black lines ($\text{P}_{\text{1,2}}$) are associated with interactions between Si adatoms and the substrate. The green (blue) line traces the dispersion of the lower Hubbard band $\text{D}_{\text{Si}}$ ($\text{D}_{\surd{3}}$), attributed to dangling bonds of Si adatoms at the interface.
        (c) Dispersion of the lower Hubbard bands $\text{D}_{\text{Si}}$ and $\text{D}_{\surd{3}}$. The lower binding energy for the Si intercalated sample is attributed to screening by the conduction electrons in graphene. Vertical lines indicate high-symmetry points of the ($\surd{3} \times \surd{3}$)$R30^{\circ}$ (blue) and ($3\times3$) (red) Brillouin zones, as depicted in (d). Refer to the main text for detailed discussion.
        (e) SPA-LEED measurement ($E=$ \qty{150}{eV}) of a Si intercalated graphene sample prepared with technique B. Selected reciprocal lattice vectors for graphene (G), SiC, and the ($\surd{3} \times \surd{3}$)$R30^{\circ}$ and ($3\times3$) superstructures are depicted as black arrows.
        (f) Line scans along the [11$\overline{\text{2}}$0] direction of SiC for G/Si/SiC (solid, green) and Si($\surd{3}$)/SiC (dashed, blue).
        (g) Line scans along the [1$\overline{\text{1}}$00] direction of SiC for G/Si/SiC (solid, green) and the Si rich ($3\times3$) reconstruction without graphene (Si($3\times3$)/SiC) (dashed, red).
        }
        \label{ARPES_LHB}
    \end{figure}

    To conduct further investigations of the surface state, EDC's in $\overline{\text{K}_\text{G}\text{\textGamma}\text{M}_\text{G}}$ direction were obtained at various values for $k_\|$ from a sample prepared using the technique B, which are displayed in \autoref{ARPES_LHB}\,(a). In \autoref{ARPES_LHB}\,(b), comparable measurements of the Si rich ($\surd{3} \times \surd{3}$)$R30^{\circ}$ reconstruction on SiC(0001) without graphene are plotted. Refer to the Methods Section for details on the preparation of the ($\surd{3} \times \surd{3}$)$R30^{\circ}$ reconstruction.

    The contribution from the SiC valence band ($\text{VB}_\text{SiC}$) is marked by orange arrows and disperses downwards from $\text{\textGamma}$ to larger momenta. In addition, both systems show a significant contribution denoted $\text{P}_\text{1,2}$ between \qty{2}{eV} and \qty{3}{eV} for values of $k_\|$\,>\,\qty{0.5}{\angstrom^{-1}}. Similar states have been observed before \cite{emtsev_r3, Northrup_1} and can be attributed to the interaction between Si adatoms and substrate. Therefore, the existence of these backbond states suggests that Si is adsorbed at the interface, in good agreement with the results discussed above.

    In both cases, the dangling bond state, labeled as $\text{D}_\text{Si}$ and $\text{D}_{\surd{3}}$, respectively, contributes to the spectra at lower binding energies within the bulk band gap. To evaluate the dispersion, the EDC's were fitted in the area of interest using a Voigt line profile. The green and blue lines indicate the trace of the extracted maxima. Due to the low intensity of the surface state it is challenging to determine the energy locations with high accuracy.

    \autoref{ARPES_LHB}\,(c) compares the dispersion of the screened dangling bond state $\text{D}_\text{Si}$ with $\text{D}_{\surd{3}}$ of the bare ($\surd{3} \times \surd{3}$)$R30^{\circ}$ surface. The latter has a plateau around $\text{\textGamma}$ and disperses downwards at higher momenta, exhibiting minima at the ($\surd{3} \times \surd{3}$)$R30^{\circ}$ Brillouin zone's high-symmetry points indicated by vertical, blue lines. Both the dispersion and the energy location are in reasonable agreement with previous studies of this surface \cite{Johansson_r3, emtsev_r3}.
    
    The graphene sample's $\text{D}_\text{Si}$ is likewise flat at $\text{\textGamma}$. On the left side, however, a maximum forms at $\text{M}_{\surd{3}}$, which differs from the minimum found for $\text{D}_{\surd{3}}$ at the same momentum. Moreover, on the right side, the minimum at \qty{0.90}{\angstrom^{-1}} cannot be explained by the ($\surd{3} \times \surd{3}$)$R30^{\circ}$ structure, since it has no high-symmetry point at this location. But, as is evident from \autoref{ARPES_LHB}\,(d), this wave vector coincides with the $\text{K}_{3 \times 3}$ point of the ($3 \times 3$) superstructure. It should be noted that all ($\surd{3} \times \surd{3}$)$R30^{\circ}$ high-symmetry points are correspondingly ($3 \times 3$) high-symmetry points. However, a ($3 \times 3$) alone is not sufficient to explain the observed dispersion. For example, this is evident when comparing the binding energy of $\text{D}_\text{Si}$ at the two $\text{K}_{3 \times 3}$ points at \qty{0.45}{\angstrom^{-1}} and \qty{0.90}{\angstrom^{-1}}, respectively, which show a difference of \qty{0.2}{eV}. Presumably, the observed dispersion is caused by superimposed photoemission from two bands with distinct periodicities. It is hypothesized that the surface is inhomogeneous with patches of both structures. In particular, the Mott states arising from the ($\surd{3} \times \surd{3}$)$R30^{\circ}$ and ($3 \times 3$) intercalation phases create this effect. \\

    The existence of these two structures at the graphene-SiC interface was investigated by electron diffraction measurements.
    High-resolution spot profile analysis - low energy electron diffraction (SPA-LEED) was applied to a Si intercalated graphene sample and subsequently compared with equivalent measurements of the Si rich ($\surd{3} \times \surd{3}$)$R30^{\circ}$ and ($3 \times 3$) superstructures without graphene. The full SPA-LEED measurements on the ($3\times3$) and ($\surd{3} \times \surd{3}$)$R30^{\circ}$ reconstructions are shown in Section 3 of the Supplementary Material. \autoref{ARPES_LHB}\,(e) shows a representative SPA-LEED measurement of a Si intercalated sample. The graphene (G) and SiC contributions are superimposed by a ($3 \times 3$) reconstruction of the SiC(0001) surface, demonstrating that at least part of the Si atoms at the graphene-SiC interface assemble in this geometry. On the other hand, no unambiguous conclusions can be drawn about the parallel existence of the ($\surd{3} \times \surd{3}$)$R30^{\circ}$ structure, since its reciprocal lattice points coincide with the ($3 \times 3$) periodicity.
    
    \autoref{ARPES_LHB}\,(f) compares the line scan along the [11$\overline{\text{2}}$0] direction of SiC from the Si intercalated sample (solid, green) with an analogous measurement of the ($\surd{3} \times \surd{3}$)$R30^{\circ}$ superstructure (dashed, blue). The former shows maxima at the same positions where peaks are observed for the ($\surd{3} \times \surd{3}$)$R30^{\circ}$ reconstruction, as well as the first-order spots of graphene, which, together with the (00) spot, are superimposed by a broad background that is absent for SiC as well as the reconstruction spots. Recent electron diffraction studies have identified this bell-shaped component as a hallmark of freestanding epitaxial graphene formation \cite{mamiyev2022sn} and a measure of uniformity, the presence of which indicates a high-quality 2D material \cite{chen2020high, petrovic2021broad, omambac2021non}. It should also be noted that next to the first order ($\surd{3} \times \surd{3}$)$R30^{\circ}$ spots of the Si intercalated sample, ($6\surd{3} \times 6\surd{3}$)$R30^{\circ}$ periodicities are observed which are attributed to residual ZLG.

    The line scan along the [1$\overline{\text{1}}$00] direction for the Si intercalated sample (solid, green) is shown in \autoref{ARPES_LHB}\,(g) together with an equivalent measurement of the Si rich ($3\times3$) reconstruction without graphene (dashed, red). The positions of both the ($3\times3$) and SiC spots are in agreement between the two systems. \\

    Finally, the key parameters of a Mott insulator, the delocalization and the on-site repulsion (Hubbard parameter) $U$ were determined from the acquired dispersions in \autoref{ARPES_LHB}\,(c). The delocalization is proportional to the band width $b$ of the unperturbed band structure. Given that the lower and upper Hubbard band are isomorphic, $b$ was calculated by doubling the band width of the LHB \cite{bechstedt2004electron, seyller_SiC_surfaces}. The band width for both the intercalated Si beneath graphene and the Si rich ($\surd{3} \times \surd{3}$)$R30^{\circ}$ structure without graphene was found to be \qty{0.4}{eV}, which is consistent with other studies of the ($\surd{3} \times \surd{3}$)$R30^{\circ}$ reconstruction \cite{Johansson_r3, emtsev_r3}.

    To calculate the Hubbard parameter — the splitting between the lower and upper Hubbard band — it is necessary to know the energy location of the UHB. Using $k$-resolved inverse photoemission, Themlin et al. \cite{Themlin} obtained a value of \qty{1.1}{eV} above the Fermi level for the bare ($\surd{3} \times \surd{3}$)$R30^{\circ}$ structure. The binding energy of the LHB at $\text{\textGamma}$ was determined to \qty{1.2}{eV} by Johansson et al. \cite{Johansson_r3}, which is consistent with the repeated measurements presented here. Therefore, the Hubbard parameter for the bare ($\surd{3} \times \surd{3}$)$R30^{\circ}$ surface adds up to \qty{2.3}{eV}.

    For Si underneath graphene the location of the UHB is unknown, such that only a lower bound for $U$ can be determined. The model proposed in Ref. \cite{emtsev_correlation_effects} was used for this purpose: The minimal separation between the Mott states corresponds to the situation where the UHB accepts electrons from the n-type substrate and pins the Fermi level.
    Accordingly, the splitting can be obtained from the lower edge of the LHB, which is separated from the Fermi level by $U$. Therefore, the Hubbard parameter for Si underneath graphene is at least 0.8\,eV.

%=================
%	3 - Conclusion 
%=================	

    \section{Conclusion}

    The present work demonstrated three pathways for the intercalation of Si below epitaxial graphene on SiC(0001). In all cases, a decoupling of the graphene layer was achieved, as evidenced by photoelectron spectroscopy.
    Moreover, a fully occupied surface state at \qty{0.6}{eV} was detected, which is attributed to the lower Hubbard band of the half-filled Si dangling bonds at the interface, that undergo a Mott-Hubbard metal-insulator transition \cite{Johansson_r3, Northrup_2, furthmüller_3x3, Hubbard, emtsev_r3}. Electron diffraction measurements identified a ($3 \times 3$) superstructure of the Si atoms at the graphene-SiC interface. However, a detailed analysis of the dispersion of the lower Hubbard band revealed no unambiguous symmetry. It is hypothesized, that the surface consists of patches with different periodicities, e.g. ($\surd{3} \times \surd{3}$)$R30^{\circ}$ and ($3 \times 3$). Nevertheless, in contrast to the analogous structure without graphene, the on-site repulsion of the dangling bonds was observed to be screened by the conduction electrons of the graphene layer, presumably shrinking the band gap of the Mott insulator.

    The ratio of on-site repulsion and delocalization is the key parameter for the development of a Mott insulator; in this case, it was determined to be $U/b \geq 2.0$. According to Ref. \cite{Hubbard3}, $(U/b)_{\text{critical}} = \surd{3}/2 \approx 0.9$ is the critical value for a Mott-Hubbard metal-insulator transition. This system provides an experimental platform to gradually reduce the Hubbard parameter by enhancing the screening through alkali doping of the graphene layer, ensuring that Mott and dopant state hybridization is avoided. Examining the Mott states as they get closer to the crucial value and finally witnessing a change to a bad-metal behavior should be the goal for further investigations.

    No hybridization of the Mott states with the Dirac bands was observed, although the existence of dangling bonds just below the graphene was established. This observation is in contrast to theoretical predictions \cite{witt2025quantum} and a recent experimental realization of a ($\surd{3} \times \surd{3}$)$R30^{\circ}$ reconstructed Sn layer beneath graphene \cite{ghosal2024electronic}, both of which found a hybridization of graphene and Mott insulator. Besides the different orientations of the dangling bonds and C atoms mentioned above, another factor to consider is the presence of the Si rich ($3 \times 3$) structure. Although it is also a Mott insulator \cite{furthmüller_3x3}, the number of dangling bonds is only one third of that of the ($\surd{3} \times \surd{3}$)$R30^{\circ}$ reconstruction \cite{schardt_3x3}, providing a lower density of states available for interaction and thus suppressing hybridization.

%=================
%	0 - Experimental
%=================	

\section{Methods}	
    
    \subsection*{Sample preparation}

    n-type 4H-SiC substrates (Cree, Inc.) were wet-chemically cleaned as described in \cite{nasschemie}. Afterwards, the photoresist AZ5214E was applied to the wafer pieces in an ultra-sonic bath for \qty{10}{min} at \qty{30}{\degreeCelsius}, according to the polymer-assisted sublimation growth technique established by Kruskopf et al. \cite{PASG}. The sample surface was then rinsed for \qty{1}{min} with isopropanol and dried on a spin coater at \qty{6000}{rpm} for \qty{1}{min} under a constant N$_\text{2}$ flow. The ZLG formation was carried out in an inductively heated hot wall reactor \cite{Ostler}: The samples were degassed at \qty{400}{\degreeCelsius} for \qty{10}{min} and annealed at \qty{900}{\degreeCelsius} for \qty{30}{min}. After cooling back to room temperature, two heating steps at \qty{1200}{\degreeCelsius} and \qty{1450}{\degreeCelsius} for \qty{12}{min} and \qty{6}{min} were performed under Ar atmosphere.
    
    Si deposition was executed in UHV (base pressure < \qty{5e-10}{mbar}) using an EFM3 rod evaporator from Focus. The evaporation rate was kept constant for all experiments. The ion flux was tracked with the attached flux monitor and set to \qty{2}{nA}. The attenuation of the XP spectra, assuming \qty{26}{\angstrom} \cite{Diss_Emtsev} for the mean free path of the photoelectrons, allowed the determination of the thickness of the deposited material. The sample temperature during annealing was tracked with an infrared pyrometer (emissivity 0.9). To achieve hydrogen intercalation, the ZLG was exposed to ultra-pure hydrogen for \qty{90}{min} at \qty{550}{\degreeCelsius} through annealing in a dedicated contactless infrared heating system \cite{diss_sieber}. The pressure (flow rate) was set to \qty{880}{mbar} (\qty{0.9}{slm}).

    The ($3\times3$) structure was prepared by exposing 4H-SiC wafer pieces to a Si flux for \qty{60}{min} while heating at \qty{750}{\degreeCelsius}. The ($\surd{3} \times \surd{3}$)$R30^{\circ}$ reconstruction was then obtained by annealing at \qty{1000}{\degreeCelsius} for \qty{20}{min} \cite{SiC_starke}.
  
    \subsection*{Photoelectron spectroscopy}	
	 
    X-ray photoelectron spectroscopy was performed with Al-K$_{\text{\textalpha}}$ radiation (\qty{1486.6}{eV}) generated by a Specs XR50M X-ray source and monochromatized with a Specs Focus 500 monochromator. After brief transport through air to another UHV system (base pressure < \qty{2e-10}{mbar}) and subsequent degassing of the sample at \qty{500}{\degreeCelsius}, angle-resolved measurements were done with HeI radiation (\qty{21.2}{eV}) from a Specs UVS 300 monochromatized with a Specs TMM304. While \qty{20}{eV} was used for the Dirac cone measurements, \qty{10}{eV} pass energy was employed to measure the core levels and energy-momentum maps. Photoelectrons were detected with a 2D CCD or channeltron detector attached to a Specs Phoibos 150 hemispherical analyzer. The core level measurements have a total energy resolution of \qty{100}{meV}. The energy resolution for the Dirac cone measurements and energy-momentum maps is \qty{40}{meV} and \qty{30}{meV}, respectively, because of the different pass energies.

    \subsection*{Spot profile analysis - low energy electron diffraction}
    For SPA-LEED measurements, the Si intercalated graphene samples were transferred into the dedicated chamber using a UHV suitcase and then degassed at \qty{500}{\degreeCelsius}. The Si rich reconstructions on SiC(0001) without graphene were capped with additional Si after preparation. Transport to the SPA-LEED chamber was through air. Before the measurements, the surface reconstructions were restored by annealing at \qty{750}{\degreeCelsius} for the ($3\times3$) and \qty{1050}{\degreeCelsius} for the ($\surd{3} \times \surd{3}$)$R30^{\circ}$. A SPA-LEED setup \cite{scheithauer1986new} with \qty{200}{nm} transfer width for surface studies was used to control and study both the Si intercalation process and the interface structures.

%=================
%	0 - Author contributions
%=================	

    \section*{Author contributions}

    C.T.\ and T.S.\ conceived the project. 
    The sample preparation was carried out by N.T.
    The PES measurements were performed and analyzed by N.T.
    The SPA-LEED measurements were performed by Z.M. and analyzed by N.T., and Z.M.
    All authors discussed the results. N.T.\ made the figures, and wrote the paper, with significant input from all authors.

%=================
%	0 - Acknowledgments
%=================	

    \section*{Acknowledgments}	
    
    This work was supported by the DFG within the Research Unit FOR5242 (project 449119662).

%=================
%	0 - Data Availablity
%=================	

    \section*{Data Availability}	
    
    The data that support the plots within this paper are available from the corresponding author upon reasonable request.

%=================
%	0 - Conflict of Interest
%=================	

    \section*{Conflict of Interest}	
    
    The authors declare no conflicts of interest.

%=================
%	Literaturverzeichnis
%=================	

\end{document}

\maketitle

%=================
%	C1s 
%=================	

    \section{C\,1s Core Level Spectra}

    \begin{figure}[b!]
        \centering
        \includegraphics[scale = 1]{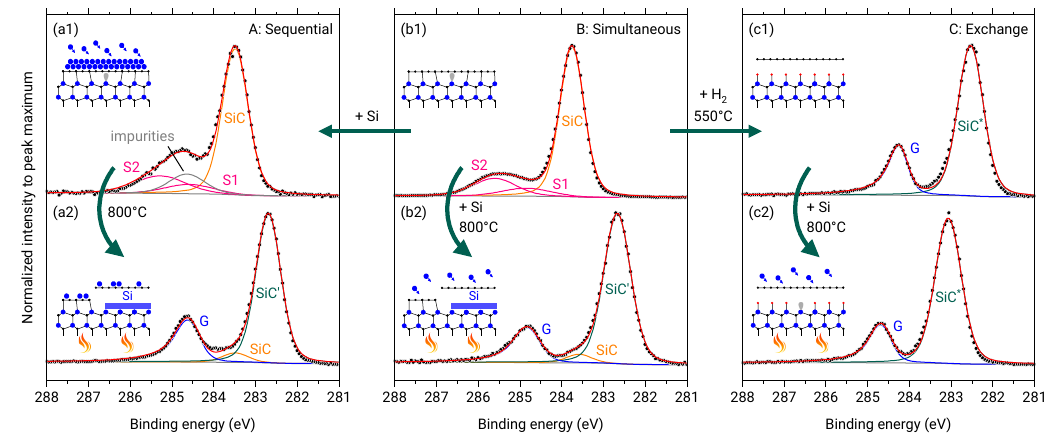}
        \caption{X-ray photoelectron spectra of the C\,1s core level, recorded after different preparation steps, all starting from an epitaxially grown zero layer graphene (ZLG) sample. A representative spectrum for the ZLG is shown in (b1), with contributions from bulk C (SiC) and ZLG atoms (S1 and S2).
        Method A: Sequential. (a1) After Si deposition, the C\,1s spectrum shows an additional component due to impurities from the source. (a2) Annealing the sample to \qty{800}{\degreeCelsius} causes significant spectral changes: (i) A new bulk component (SiC') appears at lower binding energies, indicating a modification of the interface. (ii) The ZLG components are replaced by an asymmetric peak (G) attributed to the metallic QFMLG.
        Method B: Simultaneous. (b2) Si deposition and annealing were performed simultaneously, giving a spectrum similar to (a2).
        Method C: Exchange. (c1) The ZLG was decoupled via H intercalation by annealing in a hydrogen environment, resulting in a spectrum with a shifted bulk contribution (SiC$^{*}$) and a graphene (G) component, reflecting hydrogenation of the graphene-SiC interface. (c2) Subsequent Si deposition and annealing resulted in an energy shift of both peaks.
        The insets show ball-and-stick models illustrating the individual preparation steps, with Si, C, and H atoms color-coded in blue, black, and red, respectively.
        }
        \label{C1s}
    \end{figure}

    In addition to the Si\,2p spectra discussed in Section 2.1 of the main text, corresponding C\,1s spectra were recorded for each preparation step. \autoref{C1s}\,(b1) shows a representative spectrum of a ZLG sample, which was the starting point for all preparation approaches. The spectrum has been fitted with a bulk C (SiC) and two ZLG components (S1 and S2), where S1 represents the ZLG C atoms coupled to bulk Si atoms. S2 is the contribution from C atoms that are exclusively bound within the ZLG, causing a chemical shift to higher binding energies compared to S1. The intensity ratio S2:S1 was constrained to 2:1, and the energy split between SiC and S1 was set to \qty{1.1}{eV} \cite{Sublimationswachstum}.

    After Si deposition at room temperature, a new component had to be introduced in the C\,1s spectrum (see \autoref{C1s}\,(a1)), which is attributed to impurities from the source. \autoref{C1s}\,(a2) shows the C\,1s spectrum after subsequent annealing to \qty{800}{\degreeCelsius}, which leads to several changes in the spectral shape: (i) The impurities have disappeared, indicating that they desorbed during annealing. (ii) A new bulk component (SiC') appears at lower binding energies, showing, in analogy to the Si\,2p spectra, a change in band bending due to interfacial modifications by intercalated Si. (iii) The ZLG components S1 and S2 are replaced by an asymmetric peak (G) associated with the metallic QFMLG formed by intercalation. A Mahan line shape was used to fit the graphene contribution.

    The second preparation approach, in which Si was deposited on the hot ZLG sample, yielded a C\,1s spectrum (\autoref{C1s}\,(b2)) of similar shape to \autoref{C1s}\,(a2), indicating a similar interfacial transformation, consistent with the results discussed in the main text.

    For the third intercalation technique, exchange intercalation, the ZLG was first decoupled by hydrogen. \autoref{C1s}\,(c1) shows the C\,1s core level spectrum, which consists of contributions from the hydrogen-terminated bulk (SiC$^{*}$) and the QFMLG (G) on top of it. Through Si deposition and simultaneous annealing (\autoref{C1s}\,(c2)) a shift to higher binding energies is observed for both components, clearly showing a change in SiC band bending and graphene doping as discussed in the main text. Note also that the spectrum in \autoref{C1s}\,(c2) can still be fitted with two components, verifying the absence of deintercalation during this process.

%=================
%	Continued Exchange Intercalation 
%=================	

    \section{Continued Exchange Intercalation}

    \begin{figure}[b!]
        \centering
        \includegraphics[scale = 1]{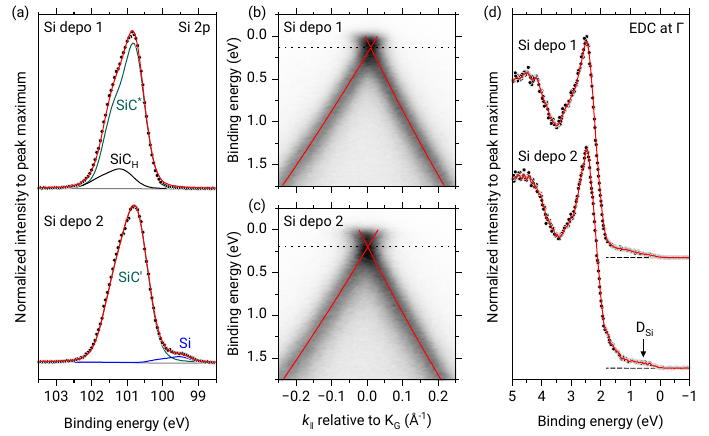}
        \caption{Electronic structure of the sample prepared by the exchange intercalation approach after different preparation cycles, labeled Si depo 1 and 2, where Si depo 1 corresponds to the stage discussed in the main text.
        (a) Si\,2p core level spectra. Both spectra show a bulk Si contribution, denoted as SiC$^{*}$ and SiC', which is attributed to H and Si passivated SiC, respectively. SiC$_\text{H}$ takes into account the chemical shift resulting from Si-H coupling. Only the spectrum after Si depo 2 shows a shoulder on the low binding energy site attributed to elemental Si at the graphene-SiC interface.
        (b)-(c) Energy-momentum maps in the vicinity of the graphene $\text{K}_{\text{G}}$ point in the direction perpendicular to $\overline{\text{\textGamma}\text{K}_\text{G}}$. After Si depo 2 (c) the graphene shows a stronger n-doping than after Si depo 1 (b). The red lines correspond to the tight-binding band structure of graphene in the nearest-neighbor approximation fitted to the maxima of momentum distribution curves. The horizontal dashed lines indicate the Dirac energy.
        (d) Energy distribution curves at \textGamma. Only the spectrum after Si depo 2 shows the surface state labeled $\text{D}_{\text{Si}}$.
        }
        \label{Si_depo_2}
    \end{figure}

    As demonstrated in the main text, the exchange intercalation approach, where Si is intended to replace H in a sample previously intercalated with hydrogen, gave results clearly different from the other two techniques.
    %This is attributed to the formation of a Si-H compound at the graphene-SiC interface. In this section, it is shown that by further processing the sample, similar results can be obtained by exchange intercalation.
    The changes observed after the first preparation cycle are attributed to marginal H desorption and the resulting dangling bonds.

    \autoref{Si_depo_2} compares the results for two subsequent Si deposition and annealing cycles, labeled Si depo 1 and 2, where Si depo 1 corresponds to the results presented in the main text. Note that Si depo 1 lasted \qty{30}{min}, while Si depo 2 took another \qty{90}{min}. \autoref{Si_depo_2}\,(a) shows the two corresponding Si\,2p core level spectra.
    Only the Si\,2p after Si depo 2 shows a shoulder at the lower binding energy site, which is attributed to elemental Si at the graphene-SiC interface. It is assumed that further processing of the sample increased H desorption, allowing Si to intercalate. Consequently, the bulk contribution SiC$^{*}$ (SiC') after Si depo 1 (Si depo 2) is attributed to H (Si) passivated SiC.
    %Besides the Si bulk contribution (SiC'), which is present in both spectra. Note also that the Si\,2p spectrum after Si depo 2 shows no evidence of deintercalation.
    \autoref{Si_depo_2}\,(b) and (c) show the graphene Dirac cones after the two preparation cycles. As can be seen from the superimposed fit (red lines), the n-doping increased with further processing of the sample. The doping was doubled from \qty{1e12}{cm^{-2}} in (b) to \qty{2e12}{cm^{-2}} in (c), approaching the doping concentrations obtained for the other two intercalation techniques A and B.
    \autoref{Si_depo_2}\,(d) shows the EDC's at \textGamma\ after both preparation steps.
    %The EDC after Si depo 2 shows a faint peak labeled $\text{D}_{\text{Si}}$ within the bulk band gap that is completely absent in the EDC after Si depo 1.
    As mentioned in the main text, the EDC after Si depo 1 shows an increased background in the bulk band gap compared to a pure H intercalated sample, indicating the presence of isolated dangling bonds. Upon further processing, this background evolves into a more peak-like feature as seen in the EDC after Si depo 2, indicating the formation of the lower Hubbard band and consequently the emergence of a Mott insulator at the graphene-SiC interface.

    In conclusion, it has been shown that further processing of the exchange intercalation can bring the sample closer to the state achieved by the preparation approaches A and B. The evidence for this trend is multiple: (i) By further processing, a shoulder appears in the Si\,2p spectrum, indicating a gradual desorption of H and an accumulation of elemental Si at the graphene-SiC interface. (ii) The graphene doping increases and approaches the doping concentration obtained by the techniques A and B. (iii) With the presence of elemental Si at the graphene-SiC interface, a fully occupied surface state occurs within the bulk band gap, which, in analogy to the results from the other intercalation methods, is attributed to the lower Hubbard band of Si dangling bonds.

%=================
%	SPA-LEED 3x3 and r3 
%=================	

    \section{SPA-LEED of Si rich Reconstructions on SiC(0001)}

\begin{figure}[h]
        \centering
        \includegraphics[scale = 1]{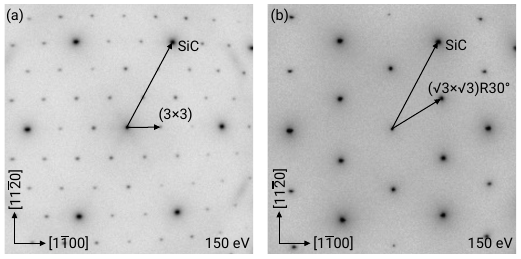}
        \caption{SPA-LEED measurements ($E=$ \qty{150}{eV}) of the Si rich ($3\times3$) (a) and ($\surd{3} \times \surd{3}$)$R30^{\circ}$ (b) reconstruction on SiC(0001). Black arrows indicate selected reciprocal lattice vectors of SiC and the reconstructions, respectively.
        }
        \label{SPALEED}
    \end{figure}

    For comparison with the Si intercalated graphene samples, the Si rich ($3\times3$) and ($\surd{3} \times \surd{3}$)$R30^{\circ}$ reconstructions were prepared on SiC(0001). Refer to the Methods Section in the main text for details on the preparation. \autoref{SPALEED} shows the corresponding SPA-LEED measurements of both structures, clearly demonstrating the presence of the aforementioned superstructures superimposed on the SiC diffraction pattern. The line scans shown in Figure 3 of the main text were measured along the high-symmetry directions [11$\overline{\text{2}}$0] and [1$\overline{\text{1}}$00] as indicated in \autoref{SPALEED}.

